\documentclass[prd,twocolumn,showpacs,amsmath,amssymb]{revtex4}
\usepackage{graphicx}

\newcommand{\dd}{{\text{d}}}

\newcommand{\ctime}{{\mathcal{C}}_{\text{time}}}
\newcommand{\dreff}{\Delta R_{\text{eff}}}

\begin{document}
\title{Long-lived oscillons from asymmetric bubbles: existence and stability}
\author{Artur B. Adib}
\email{artur.adib@brown.edu}
\altaffiliation[Address after September 2002:]{
Box 1843, Brown University,
Providence, RI 02912, USA.}

\author{Marcelo Gleiser}
\email{gleiser@logos.dartmouth.edu}
\affiliation{Department of Physics and Astronomy, Dartmouth College,
Hanover, NH 03755, USA}

\author{Carlos A. S. Almeida} \email{carlos@fisica.ufc.br}
\affiliation{Departamento de F\'{\i}sica, Universidade Federal do
Cear\'{a}, Caixa Postal 6030,  60455-760, Fortaleza, Cear\'{a}, Brazil}

\date{\today}

\begin{abstract}
The possibility that extremely long-lived, time-dependent, and localized
field configurations (``oscillons'') arise during the collapse of
{\it asymmetrical} bubbles in 2+1 dimensional $\phi^4$ models is
investigated.
It is found that oscillons can develop from a large spectrum of elliptically
deformed bubbles. Moreover, we provide numerical evidence that such oscillons
are: a) circularly symmetric; and b) linearly stable against small
arbitrary radial and angular perturbations. The latter is based on a
dynamical approach designed to investigate
the stability of nonintegrable time-dependent
configurations that is capable of probing
slowly-growing instabilities not seen through the usual ``spectral'' method.
\end{abstract}


\maketitle


\section{Introduction}

The existence of classical field configurations exhibiting soliton-like
properties is not only an interesting consequence of nonlinear effects in
field theory, but also an important ingredient in the understanding of
nonperturbative effects in particle physics \cite{rajaraman}. Apart from
one-dimensional kinks, examples of these structures in 3+1 dimensions
include nontopological solitons (NTS) \cite{lee} and $Q$ balls \cite{coleman}:
in both cases, choosing a simple harmonic time-dependence for the scalar field
allows one to obtain a static solution of the field equations, describing a
spherically-symmetric configuration which, for a range of parameters, may be the
lowest energy state. Such configurations may be found in extensions of the
standard model, supersymmetric or not, as has been suggested recently
\cite{nts-obs}. They may be sufficiently stable as to allow for a quantization
procedure and form a legitimate bound state (see e.g. \cite{lee}).
NTSs and Q-balls have also been of great interest to applications
of particle physics to the early universe, often being proposed as possible
candidates for dark matter \cite{nontop-cosm}. A time-dependent, long-lived, and
localized configuration in 3+1 dimensional $\phi^4$ scalar field theory was
re-discovered and thoroughly studied by one of us and collaborators some years
ago \cite{osc1,osc2}. It was shown that these configurations, named in
Ref. \cite{osc1} ``oscillons,'' naturally arise from collapsing unstable
spherically-symmetric bubbles in models with symmetric and asymmetric
double-well potentials, being mainly characterized by a rapid oscillation of the
field at the bubble's core. Their relevant feature is that, albeit not strictly periodic,
they possess a {\em very} long lifetime, of order $10^3-10^4 m^{-1}$
($\hbar=c=1$ in this work). Oscillons may be thought of as the
higher-dimensional cousins of one-dimensional breather states found from
kink-antikink bound states \cite{breathers}. Just as kink-antikink pairs may be
thermally or quantum-mechanically nucleated through nonperturbative processes, so
may oscillon states, although here the calculation must be done in real and
not Euclidean time. More recently, Gleiser and Sornborger investigated
whether oscillons are present in 2+1 dimensions, finding not only that they do
exist, but also that their lifetime is {\em at least} of order $10^7 m^{-1}$
\cite{sornb}.
Motivated by this result, in the present paper we investigate two
important related questions: first, if oscillons still appear during the
collapse of {\it asymmetric} -- as opposed to symmetric -- initial
configurations; and, second, if they are stable against small angular and
radial fluctuations. Since this implies that we will be dealing with a
higher-dimensional parameter space, we restrict ourselves here to 2+1
dimensional oscillons. Apart from being of interest in their own right, we
expect that our results will be indicative of the behavior of 3+1 dimensional
oscillons. We also note that it should be quite easy to build oscillons from
more complicated field theories, including interactions between the ``oscillon''
field and other scalar or fermionic fields. The robustness of these
configurations, as demonstrated here, should provide enough motivation for a
careful search of such generalized oscillons (and possibly more realistic) in
the near future.

The paper is organized as follows. By means of a numerical scheme suited to
tackle long-lived configurations (described in the Appendix), in Sec.
\ref{dynamics} we show that oscillons quickly appear during the collapse of
most elliptically deformed bubbles and, moreover, that they are {\em all}
circularly symmetric and extremely long-lived, leaving no trace of the initial
asymmetry. This suggests that oscillons can be understood as attractors in
field-configuration space, ordered spatio-temporal structures that emerge during
the nonlinear evolution of a wide variety of initial configurations. In fact, in
2+1 dimensions, the attractor basin is quite deep, as was initially hinted in
Ref. \cite{sornb} and will be further shown here. We then move on to study,
through a dynamical approach, whether these symmetric configurations are
stable against small asymmetric perturbations, finding no indication of
spectral instability (Sec.
\ref{stab}). We conclude in Sec. \ref{concl} summarizing our results and
pointing out future avenues of research.


\section{Oscillons from asymmetrical bubbles}

\label{dynamics}

The Lagrangian density for our 2+1 dimensional scalar field theory is:
\begin{equation}
\label{lagrangian}
{\cal L} = \frac{1}{2} (\partial_{\mu} \phi) (\partial^{\mu} \phi) -
\frac{\lambda}{4}               \left( \phi^2 - \frac{m^2}{\lambda} \right)^2,
\end{equation}
with $\mu=0,1,2$. We introduce dimensionless variables by rescaling the
coordinates and the field as
$x'_\mu = x_\mu m$ and $\phi' = \phi \sqrt{\lambda} / m$
(henceforth we drop the primes). The energy and the equation of motion are
\begin{equation}
\label{energy}
  E [\phi]= \frac{m^2}{\lambda}\int \dd^2 x \left[ \frac{1}{2}
(\partial \phi/ \partial t)^2 + \frac{1}{2} (\nabla \phi)^2
+ \frac{1}{4} (\phi^2 - 1)^2 \right],
\end{equation}
and
\begin{equation}
  \label{eqmotion}
 \frac{\partial^2 \phi}{\partial t^2} = \nabla^2 \phi - (\phi^3 - \phi),
\end{equation}
respectively. So far, all previous studies have obtained oscillons from symmetric
initial configurations,
with either thick or thin walls
(Gaussian or tanh profiles, respectively). We will restrict our investigation
to Gaussian initial profiles, as these proved to be the most interesting in
2+1 dimensions (cf. \cite{sornb}). For convenience, we will follow Ref.
\cite{sornb} and restrict the initial field
configuration to interpolate between the two minima of the potential. Of
course, one could select different values for the initial
value of the field at the core [$\phi(t=0,r=0, \theta)$]:
as it was argued in Ref. \cite{osc2}, as long as the value
of the field at the core probes
the nonlinearity of the potential, and the initial configuration has an energy
above the ``plateau'' energy (the energy of the oscillon configuration),
oscillons are bound to appear.

\begin{figure}
\includegraphics[width=200pt]{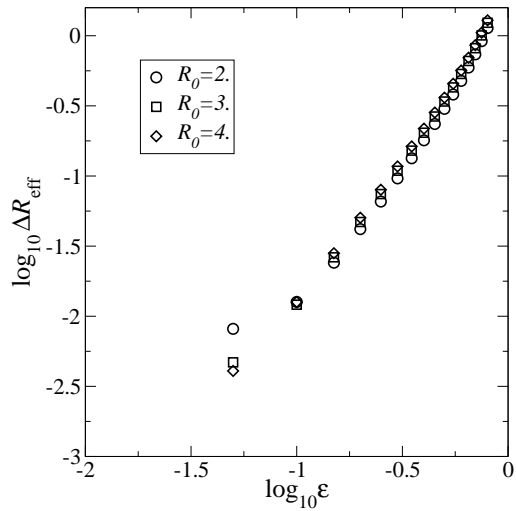}
\caption{
  \label{fig:DeltaReff}   Logarithm of the ``effective radial dispersion''
  $\dreff(t=0)$ vs.  eccentricity $\varepsilon$ for the ansatz
(\ref{asym_ansatz}).
$\dreff$ is clearly a good measure of asymmetry, increasing always monotonically
with $\varepsilon$.}
\end{figure}

The asymmetry in the initial field configuration is introduced by means
of an elliptical deformation:
\begin{equation}
\label{asym_ansatz}
  \phi(\rho,\theta,0)= -2 \exp \left[ -\frac{r^2(\rho,\theta)}{R_0^2} \right] +
 1,
\end{equation}
where:
\begin{equation}
  r^2(\rho,\theta)=\frac{\rho^2}{1-\varepsilon^2 \cos^2 \theta},
\end{equation}
with $\varepsilon:[0,1)$ the bubble eccentricity, $R_0$ the bubble ``radius''
and $\rho,\theta$
polar coordinates (notice that this expression reduces to the usual symmetric
ansatz when $\varepsilon=0$).
We note in passing that a similar parameterization was adopted in
the study of eccentric pulsons in the sine-Gordon theory \cite{chris}.

\begin{figure}
\includegraphics[width=200pt]{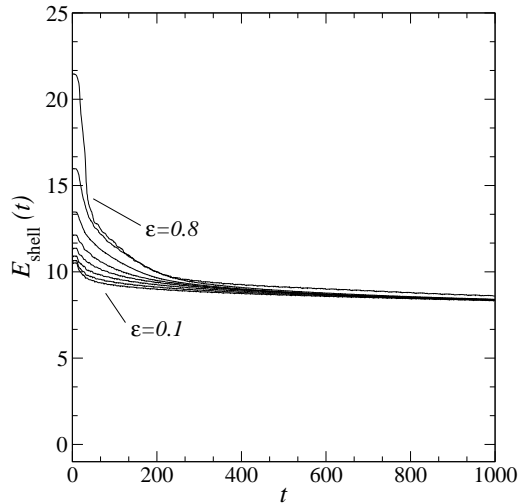}
\caption{ \label{fig:r02}
    Oscillon energy in a shell of radius $R_s=5R_0$ for $R_0=2.0$ and
    $0.1\le \varepsilon \le 0.8$, from bottom to top (only integer multiples
    of $\varepsilon=0.1$ are shown).}
\end{figure}

\begin{figure}
\includegraphics[width=200pt]{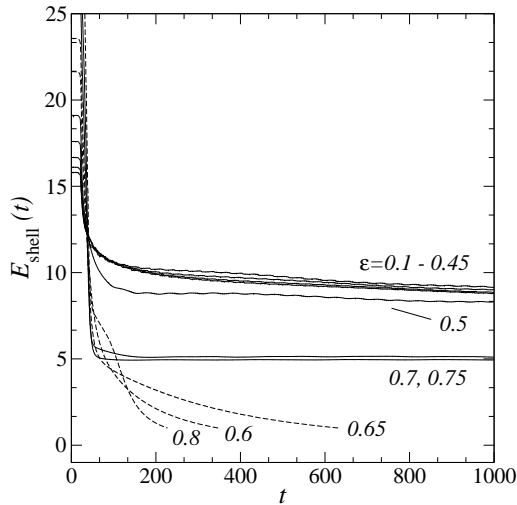}
\caption{  \label{fig:r03}
Oscillon energy in a shell of radius $R_s=5R_0$, now for $R_0=3.0$ and
$0.1\le \varepsilon \le 0.8$, with steps $\Delta \varepsilon=0.05$.
Dashed lines indicate unstable configurations.}
\end{figure}

In order to {\em measure} the asymmetry of the field configuration, we take
advantage of the effective radius $R_{\text{eff}}$ defined in Ref. \cite{osc2}
and introduce an ``effective radial dispersion'' $\Delta R_{\text{eff}}$
(see below). The effective radius could be recast in the form:
\begin{equation}
  R_{\text{eff}}(t) \equiv \langle R(\theta,t) \rangle_{\theta} ,
\end{equation}
where $\langle \cdot \rangle_{\theta} = (2\pi)^{-1} \int_0^{2\pi} \dd \theta$ is
an angular average and $R(\theta,t)$ is a $\theta$-dependent effective
radius defined by (compare Eq. (28) of Ref. \cite{osc2}):
\begin{equation}
  R(\theta,t) =
 \frac{ \int_0^{R_{\text{s}}} \dd \rho ~ \rho^2 \left[
 \frac{1}{2}(\nabla \phi)^{2} + V(\phi) \right] }
    {(2\pi)^{-1} \int_0^{2\pi} \dd \theta \int_0^{R_{\text{s}}} \dd \rho ~ \rho
 \left[ \frac{1}{2}(\nabla \phi)^{2} + V(\phi) \right] } ,
\end{equation}
with $R_{\text{s}}$ the ``shell radius'' (i.e. the radius within which we
integrate all the
quantities relative to the oscillon). This notation allows us to
construct angular averages
and dispersions in a manner analogous to time averages. We therefore define
the effective radial
dispersion as the relative root-mean-square deviation from the average radius:
\begin{equation}
  \dreff (t) \equiv \frac{ \sqrt{ \langle R^2 (\theta,t)
    \rangle_{\theta} - \langle R(\theta,t) \rangle_{\theta}^2 } }
    { \langle R(\theta,t) \rangle_{\theta} }.
\end{equation}
As shown in Figure \ref{fig:DeltaReff}, the above quantity is indeed a good
 measure of asymmetry, i.e. it increases monotonically with $\varepsilon$. It is
 approximately independent of the bubble size, being the limiting case
$\dreff = 0$ an indication of a symmetrical
state (though not exactly zero on a lattice due to its finite resolution).
We now turn to the presentation of the main numerical experiments obtained
by solving
Eq. (\ref{eqmotion}) with the initially asymmetrical bubbles (\ref{asym_ansatz})
for eccentricities $\varepsilon$ ranging from $0.1$ to $0.8$ and $R_0$ from
$2.0$ to $5.0$. This investigation can easily be extended to
greater values of $R_0$, although
this would require much longer computational times without generating results of much
physical interest.
[The computational time $\ctime$ is proportional to $L^2$, which in turn is
proportional to $(R_0)^2$, see Appendix].

\begin{figure}
\includegraphics[width=200pt]{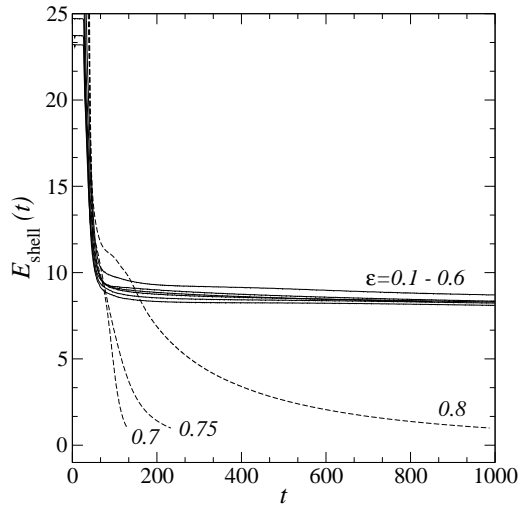}
\caption{
\label{fig:r04}
Oscillon energy for $R_0=4.0$ and different $\varepsilon$, with steps
    $\Delta \varepsilon=0.05$ (dashed lines are for unstable configurations).}
\end{figure}

\begin{figure}
\includegraphics[width=200pt]{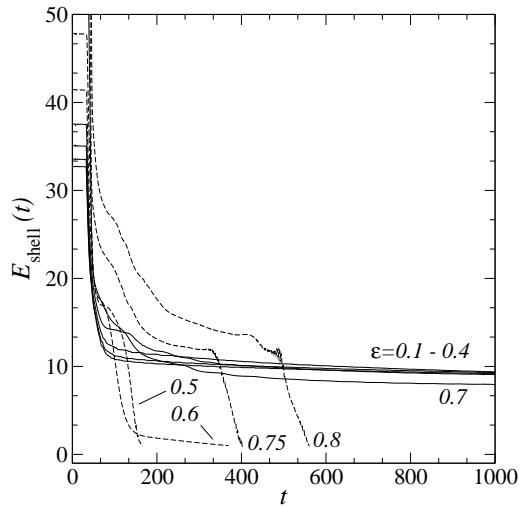}
\caption{ \label{fig:r05}    Shell energy for bubbles of
radius $R_0=5.0$ and steps $\Delta
 \varepsilon=0.05$ (dashed lines for unstable cases).}
\end{figure}

Figures \ref{fig:r02}-\ref{fig:r05} show the time evolution of the total energy within
a shell of radius $R_{\text{s}}=5R_0$ surrounding the initial configuration for different
values of $R_0$ and $\varepsilon$. It is seen that, in general, initially
asymmetric bubbles tend to decay into coherent field configurations with an
approximately constant energy plateau similar to those found in Ref.
\cite{sornb}, which focused on the evolution of symmetric configurations.
With the help of the effective radial dispersion
$\dreff$, we can investigate
whether these
configurations correspond to ``excited'' states of an oscillon
(i.e., non-spherically
symmetric configurations analogous to an excited state of a hydrogen atom
for $\ell \neq 0$)
or if the bubble asymmetry is completely lost and the system decays
into a ``ground''
(i.e. symmetric) state.

\begin{figure}
\includegraphics[width=200pt]{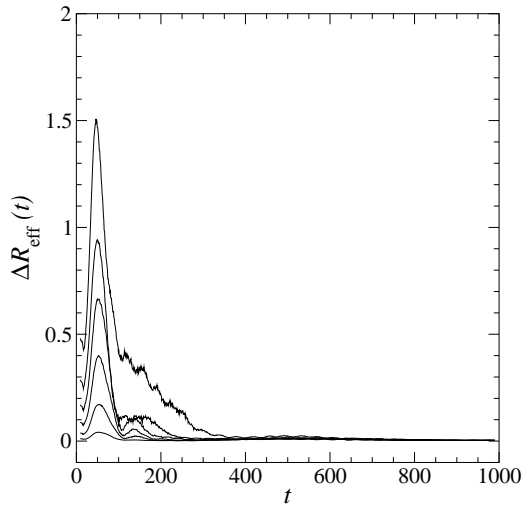}
\caption{
  \label{fig:dreff_r04}
$\dreff$ for $R_0=4.0$ and $0.1 \leq \varepsilon \leq 0.6$, from bottom to top
 (only integer multiples of $\varepsilon=0.1$ are shown). Notice how the
initial asymmetric bubble decays into a circularly symmetric configuration.}
\end{figure}

\begin{figure}
\includegraphics[width=200pt]{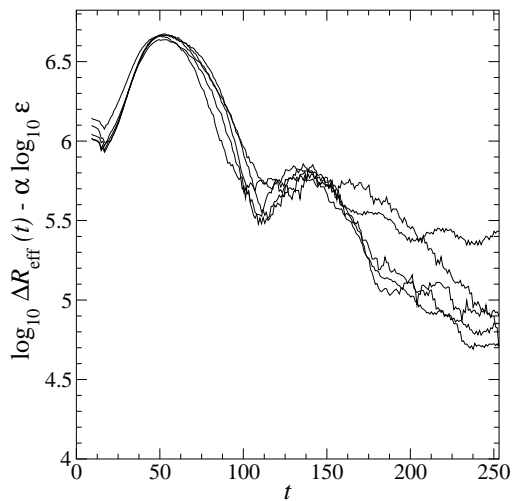}
\caption{
  \label{fig:scaling}
    Collapse of $\dreff$ in Figure \ref{fig:dreff_r04} using
    $\dreff(\varepsilon,t)=\varepsilon^{\alpha} f(t)$ for $\alpha \approx 2.05$.
Note that the time domain is restricted to $t \leq 250$, since $\dreff
\rightarrow 0$	for $t \gtrsim 250$ (causing the log scale to be ill behaved).}
\end{figure}

In Fig. \ref{fig:dreff_r04} we show the time evolution of $\dreff$ for the
 $R_0=4.0$ case.
It is clearly seen that the initially asymmetric bubble decays into a
 $\dreff=0$, symmetric
configuration, after a brief asymmetric pulsation. A similar pattern was
 observed for
all initial radii investigated here, suggesting that whenever an oscillon
stage is set, the resulting configuration is circularly symmetric. The peaks
in this figure also suggest that $\dreff$ might follow a scaling law for
 different values of $\varepsilon$. Indeed, a collapse of these curves using the
scaling
$\dreff(\varepsilon,t)=\varepsilon^{\alpha} f(t),$ where $\alpha$ is a real
constant and $f(t)$ is a function of time only, is shown in
Figure \ref{fig:scaling}.

\begin{figure}
\includegraphics[width=200pt]{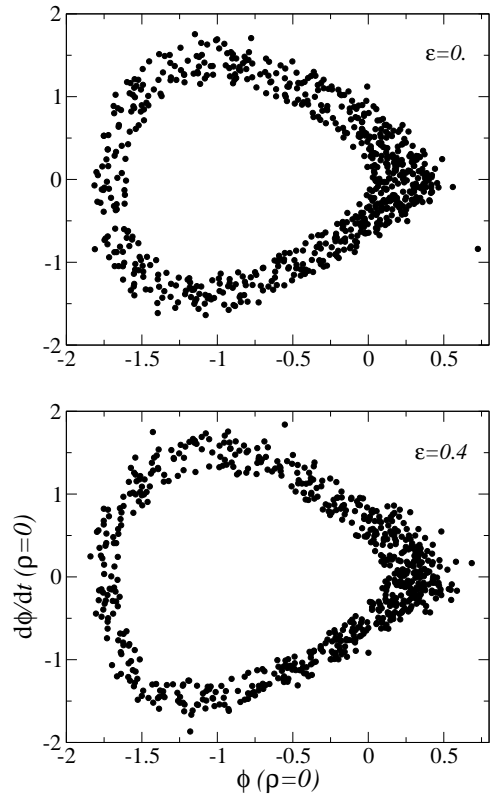}
\caption{
  \label{fig:bubcore}
    Phase portrait between $t=0$ and $t=1000$ at $\rho=0$ of a bubble entering
 an oscillon stage for $R_0=2.0$ and $\varepsilon=0$ (top) and $\varepsilon=0.4$
 (bottom): the symmetric configuration to which asymmetric bubbles decay
{\em is} an oscillon.}
\end{figure}

As the reader must have noticed, an intriguing feature of these results
is the presence of some ``instability windows'' for some values of $\varepsilon$.
 These can be observed here in the cases $R_0=3.0$ and $R_0=5.0$. Thus,
oscillons do not always appear as the asymmetric
configurations decays. A finer investigation of the parameter space for
the elliptical deformations, generalizing what
was done in great detail for spherically-symmetric 3+1 dimensional
oscillons \cite{honda},
will quite possibly reveal a very rich and detailed substructure of stable
and unstable windows. It is important to stress that once
$\Delta R_{\rm eff} \rightarrow 0$ (cf. fig. \ref{fig:dreff_r04}), the field
does settle into an oscillon, as the phase space portrait of
Fig. \ref{fig:bubcore} exemplifies. This justifies our earlier claim that
oscillons are attractors in field-configuration space.

A crucial step not yet studied is the stability analysis against small but
arbitrary asymmetric perturbations. The oscillon stability with regard to these
perturbations is
fundamental for the computation of quantum corrections around the classical
solution \cite{rajaraman,lee}
and is the subject of the next session.


\section{Stability analysis}
  \label{stab}
Our task now is to investigate whether symmetric oscillons are stable against small
radial and angular perturbations $\delta \phi(\rho,\theta,t)$, i.e. to probe
the ``linear stability'' of the oscillon. We are unaware of any previous study
where the stability of a field configuration with a time-dependent amplitude has been
tested against small fluctuations. So far, stability investigations have been
restricted to either time-independent
configurations, such as bounce solutions \cite{coleman2}, or to
configurations with a linear time-dependence in the {\em phase}, such as $Q$-balls
\cite{coleman}). The stability analysis in these two situations is
greatly simplified by the fact that the dynamical equation dictating the behavior
of the fluctuations is separable into its spatial and temporal parts; the resulting
problem reduces to finding the eigenvalues $\omega_n^2$ of a
time-independent operator (an alternative approach based on the so-called
``Bogomolnyi bound'' \cite{bogomolnyi} can also be applied in the time-independent
case, see e.g. \cite{calberto}). The existence of at least one negative eigenvalue
$\omega_n^2$ (and thus of a complex eigenfrequency $\omega_n$) signals the presence
of an exponentially-growing instability \cite{jackiw}
(a well-known example is the so-called ``bounce''
solution, which we will investigate further below \cite{coleman2}).
The present problem, however, is not
amenable to such treatment due to the anharmonic time-dependence of the oscillon; we
must consider both the space- and time-dependence of the background field,
making the stability
analysis of oscillon-type configurations
considerably more challenging both analytically and numerically,
as we will now discuss.

In order to appreciate these difficulties, let us write (in polar coordinates)
the linearized equation of motion that follows from Eq. (\ref{eqmotion})
through the substitution
$\phi(\rho,\theta,t) \rightarrow \phi_0(\rho,t) + \delta \phi(\rho,\theta,t)$,
where $\phi_0$ is the symmetric oscillon solution and $\delta \phi \ll \phi_0$
is the perturbation, i.e.,
\begin{equation}
  \label{eqlin}
  \frac{\partial^2 }{\partial t^2} \delta \phi = \frac{\partial^2}{\partial
 \rho^2} \delta \phi +
    \frac{1}{\rho} \frac{\partial}{\partial \rho} \delta \phi +
    \frac{1}{\rho^2} \frac{\partial^2}{\partial \theta^2} \delta \phi -
 (3\phi_0^2 - 1) \delta \phi.
\end{equation}
Here one might be tempted to separate the variables as $\delta \phi \equiv
R(\rho)\Theta(\theta)T(t)$. However, the resulting equations show that
one cannot get rid of the simultaneous radial and
temporal dependence of the background configuration, $\phi_0(\rho,t)$
\footnote{In Ref. \cite{osc1} it was assumed that the time-scale
of possible unstable fluctuations would be faster than typical oscillon time-scales.
However, such ``adiabatic'' assumption is too restrictive,
as the fluctuation time-scales are
{\it a priori} unknown.}.
This situation should be contrasted to the usual case where $\phi_0$ is a
time-independent solution, e.g. the bounce, or to the case where the
time-dependence of $\phi_0$ is in a phase factor
$\exp(i\omega t)$, and thus immediately eliminated in the full
equation of motion \cite{coleman}. A considerable simplification can
nevertheless be accomplished by writing
$\delta \phi \equiv \Phi(\rho,t)\Theta(\theta)$, isolating at
least the angular part of the problem. Performing such substitution gives
the pair of equations,
\begin{equation}
  \label{eqlin_rad}
  \ddot\Phi - \Phi'' - \frac{1}{\rho}\Phi' + \left( U_0(\rho,t) +
    \frac{\ell^2}{\rho^2}\right)\Phi = 0,
\end{equation}
and
\begin{equation}
  \label{eqlin_ang}
  \frac{d^2 \Theta(\theta)}{d\theta^2} = -\ell^2 \Theta(\theta),
\end{equation}
with overdots and primes indicating time and radial derivatives, respectively.
Here $U_0(\rho,t)\equiv 3\phi_0^2 -1$ and $\ell$ is a separation constant.
The solutions for $\Theta$ are trivial, viz. $\Theta \propto \exp(\pm i\ell\theta)$,
and by requiring $\delta \phi$ to be single-valued we have $\ell=0,1,2,...$.
Our original (2+1)-dimensional problem, Eq. (\ref{eqlin}), reduces therefore
to solving the above (1+1)-dimensional one, Eq. (\ref{eqlin_rad}), since
the time-dependence is present only in $\Phi$.

Our goal is to probe the linear stability of the oscillon by solving Eq.
(\ref{eqlin_rad}) for arbitrary initial conditions. The strategy is to monitor
the time evolution of the perturbations $\delta \phi$, which should grow without
bounds in the case of a linearly unstable configuration \cite{merkin,morrison}. An
obvious limitation of this approach is that it is impossible
to scan all initial values of perturbations,
viz. $\delta \phi(\rho,0)$ with $\ell=0,1,2,...$. The method thus is only indicative of
stability, not being able to provide conclusive proof. The more thorough the search,
the more one is guaranteed to show stability, at least against most types of
perturbation. This unavoidable limitation should be contrasted with the simpler case
for time-independent background configurations based on a
harmonic decomposition
$T(t)\propto \exp(i\omega_n t)$ (see e.g. \cite{coleman,jackiw}), where the existence
of exponentially unstable modes is clearly related to imaginary eigenvalues.
However, we would like to point out a limitation of the spectral method that is often
overlooked.
By restricting the analysis to an exponential
time-dependence, as in $T(t)$ above, one can obtain only
{\em spectral instabilities} of a configuration,
leaving aside other
possible forms of instabilities, for example, linear (or power-law) ones.  In other words,
a system that is spectrally stable may still be unstable against slower growing modes
\cite{morrison,lichtenberg}.
Since we are here essentially
watching the full time-dependence of $\delta \phi$, we should be able to
detect {\em any} sort of instability by observing its long-time behavior,
although in practice the infinite-time limit or a complete scan of
possible fluctuations cannot be achieved numerically.
Fortunately, we shall shortly see that typical spectral instabilities
(such as that of the bounce) do not require a long-time integration or a very wide
search, being therefore
bound to be observed through our method. Before we do so, it is worth testing
the reliability of the numerical implementation itself.

\subsection{Linear test}
The first step is to compare the numerical solution of Eq. (\ref{eqlin_rad})
with a closed-form analytical one in order to prepare and test our numerical
implementation, since the singular behavior at the origin requires a careful treatment.
This
can be done most easily by setting $U_0(\rho,t)=0$, in which case Eq. (\ref{eqlin_rad})
becomes linear and separable, with $\Phi=R(\rho)T(t)$. This gives
$T\propto \exp(\pm i\omega t)$, where $\omega$ is a separation constant, and the
equation for $R$,
\begin{equation}
  R'' + \frac{1}{\rho} R' + \left( \omega^2 - \frac{\ell^2}{\rho^2} \right) R = 0,
\end{equation}
which we recognize as Bessel's equation. By requiring regularity at the origin
and $\dot \Phi(\rho,0)=0$, the solution can be
written as
\begin{equation}
  \Phi_\ell(\rho,t) = \int_0^\infty \! \! d\omega f(\omega) J_\ell(\omega \rho)
    \cos(\omega t),
\end{equation}
where $f(\omega)$ is determined by the initial condition. Choosing
$\Phi(\rho,0)=A J_\ell(a\rho)$ we have
\begin{equation}
  f(\omega) = A \omega \int_0^\infty \! \! d\rho \, \rho J_\ell(a\rho) J_\ell(\omega\rho) = A \delta(\omega - a),
\end{equation}
and therefore
\begin{eqnarray}
  \Phi_\ell(\rho,t) & = & A \int_0^\infty \! \! d\omega \, \delta(\omega - a) J_\ell(\omega\rho) \cos(\omega t) \nonumber \\
   & = & A J_\ell(a\rho) \cos(a t).
\end{eqnarray}
The above solution maintains its shape but oscillates harmonically with
period $2\pi/a$. We have verified that our numerical implementation reproduces correctly
this analytical solution
for various values
of $a$ and $\ell$.

\begin{figure}
\includegraphics[width=200pt]{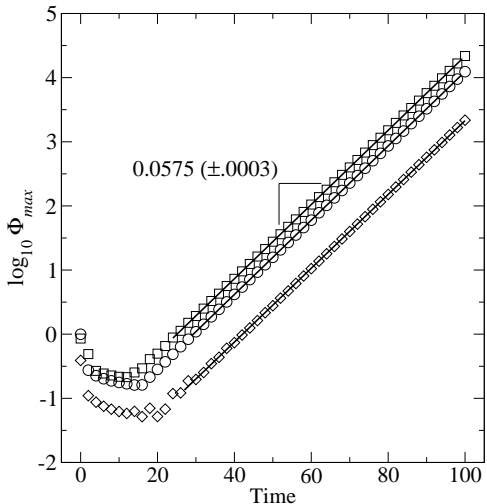}
\caption{  \label{fig:log_deltamax}
 Semi-log plot of the maximum amplitude $\Phi_{max}$ versus time
 for the bounce solution (showing here three different initial conditions).
 The slope denotes the common unstable mode eigenvalue, $\omega_n$. }
\end{figure}

\subsection{The bounce}

As a first application of our method we investigate
the stability of the so-called ``bounce'' solution \cite{coleman2}, which is
guaranteed to be spectrally unstable in any dimension greater than (1+1) due to
Derrick's theorem \cite{derrick}. (In fact, Coleman has showed
that only one negative eigenvalue exists \cite{coleman3}).
Should our method be reliable, the solution $\delta \phi$ for the case
where $\phi_0$ is a bounce solution will grow
exponentially at late times, indicating the presence of an unstable mode.

The bounce solution $\phi_b(\rho)$ is the $O(2)$-symmetric static
configuration that solves the equation
\begin{equation}
  \label{eqmotion_bounce}
 \frac{d^2 \phi}{d \rho^2} + \frac{1}{\rho} \frac{d \phi}{d \rho} = V'(\phi),
\end{equation}
with the asymmetric potential
\begin{equation}
  V(\phi)=\frac{\beta}{2} \phi^2 - \frac{1}{3} \phi^3 + \frac{1}{4}\phi^4.
\end{equation}
In order to detect the instability, we solved Eq. (\ref{eqlin_rad}) with
$U_0(\rho)=\beta - 2\phi_b(\rho) + 3\phi_b^2(\rho)$ and various initial conditions,
sweeping the lattice at every time step to find the maximum value of the perturbation, $\Phi_{max}$.
In Fig. \ref{fig:log_deltamax}, we show our results for $\beta=0.011$ [the initial conditions are
Eq. (\ref{initcond1}) with $n=0,1$ and $m=2$, and Eq. (\ref{initcond3}) for $m=2$, both with $\ell=0$];
one can clearly identify the exponential growth of $\delta \phi$
even at early times $t<100$. Also
shown is the slope of the curve, which should match the unstable eigenvalue
$\omega_n$ obtained with the usual
spectral method. [We have attempted to obtain such eigenvalue by
solving numerically the associated Schr\"odinger-like equation. However, in two spatial
dimensions the severe singularity at the origin causes a numerical
instability which we were unable to control even with sophisticated methods
\cite{liu}. Since this is not the focus of this paper, we will leave this question
aside.]

\subsection{The oscillon}
We are now ready to apply our method to the stability of the oscillon, which was obtained
here by solving Eq. (\ref{eqmotion}) with the symmetrical version ($\varepsilon=0$) of the
ansatz Eq. (\ref{asym_ansatz}). We have essentially followed the same procedure described
above for the bounce, but now evolving both $U_0$ and $\Phi$ in Eq. (\ref{eqlin_rad}).
Since
the dimensionality
of the configuration space is infinite, we chose arbitrarily the initial profiles of the
fluctuations $\Phi(\rho,t_0)$, with the only constraint that they
should vanish at $\rho\rightarrow \infty$
to ensure localization around the oscillon. The time $t_0$ was chosen to
be about $200$, since that is roughly when the initial bubbles have just decayed into
an oscillon (cf. Figs. \ref{fig:r02}-\ref{fig:r05}). Some examples of the initial
configurations investigated here are
\begin{align}
  \Phi_\ell(\rho,t_0) & = x^n \exp\left(- \frac{\rho^m}{R_0^m} \right),
\label{initcond1} \\
  \Phi_\ell(\rho,t_0) & = \sin\left(\frac{2n\pi \rho}{R_0}\right)
\exp\left( -\frac{\rho^m}{R_0^m} \right), \label{initcond2} \\
  \Phi_\ell(\rho,t_0) & = J_\ell(\rho) \exp\left( -\frac{\rho^m}{R_0^m} \right), \label{initcond3}
\end{align}
for various integer values of $n,m$ and $\ell$ (namely, $n=0,1,2$, $m=2,3$ and $\ell=0,1,2,3$).

\begin{figure}
\includegraphics[width=200pt]{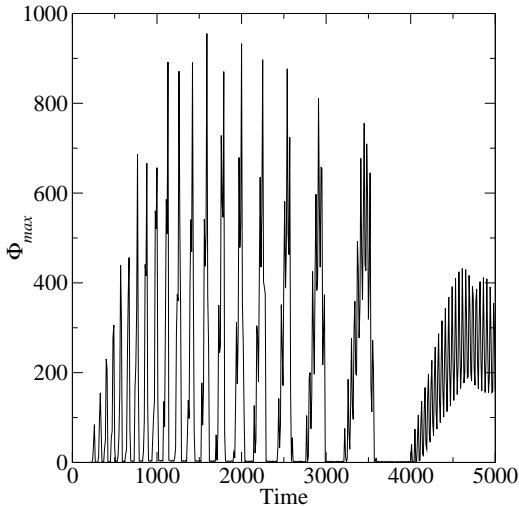}
\caption{  \label{fig:ex1}
  A typical outcome of the linear stability analysis of the oscillon, showing here
  $\Phi_{max}$ vs. time for $\ell=0$, $n=0$ and $m=2$ in Eq. (\ref{initcond1}). The
  radius of the initial bubble is $R_0=2$.}
\end{figure}

In Figure \ref{fig:ex1} we show a typical outcome of our search. In all cases
investigated we found that
the fluctuations $\Phi_{max}$ are bounded
from above, as one would expect from a linearly stable configuration.
We conclude that if, indeed, there are any unstable modes, they are sufficiently
slow-growing to justify the use of the oscillon as a stable bound state.
[We have integrated the linearized equations of motion up to $t=10^4$; see
also discussion below].
Note that the large amplitude of $\Phi$, e.g. $\sim 1000$ in Fig. \ref{fig:ex1}, does not
mean
that the condition $\delta \phi \ll \phi_0$ is violated: since the
resulting equation is {\em linear},
any solution can always be rescaled without changing its shape by choosing a
different constant prefactor for the initial conditions.

\begin{figure}
\includegraphics[width=200pt]{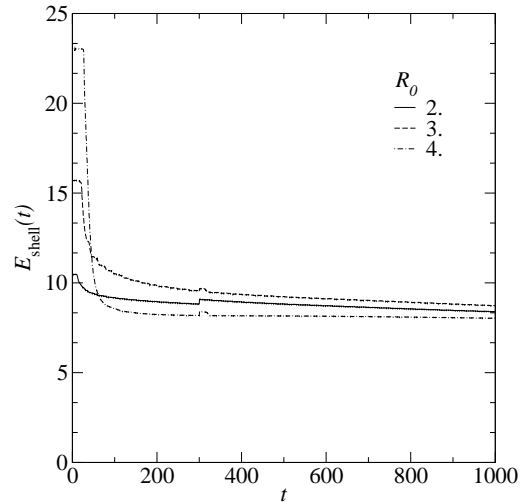}
\caption{  \label{fig:ex2}
An example of the full dynamics of the oscillon when subject to a perturbation of the
form in Eq. (\ref{initcond1}) for $n=0$ and $m=2$ at $t=300$.
The maximum amplitude of the initial perturbation - here the prefactor in Eq. (\ref{initcond1}) -
was constrained to either $0.01$ or $0.05$,
although the
results do not change appreciably even for $0.1$.
The stability is evident through the persistence
of the energy plateau. }
\end{figure}

Although not as systematic and transparent as the investigation above,
another approach to check the stability of oscillons is to superimpose the
perturbation to the full (2+1)-dimensional oscillon dynamics discussed in
Sec. \ref{dynamics}. One can then probe the oscillon stability simply by
checking the persistence of the energy plateau: if the added energy
from the perturbations is radiated away, the oscillon is stable. Due to the
dimensionality of the problem, the numerical treatment is quite more challenging than the
one use above within the linear method. Nevertheless, we have investigated the stability
of oscillons against superimposed fluctuations
for similar initial conditions. In Fig. \ref{fig:ex2} we present the outcome of a
particular choice of initial
condition for three different initial radii. The results are consistent with the previous
stability analysis, as can be seen by the persistence of the energy plateau.

On the basis of our extensive search
with many different initial conditions and long integration times,
we find it very unlikely that an
exponentially-growing mode exists. If it does, it would be either very
small and/or related to a very
``rare'' excitation; the oscillon configuration would still be stable for large times and
could be considered a legitimate (or at least a very long-lived)
bound state in semi-classical quantization.

\section{Conclusions}
  \label{concl}

We have investigated, in 2+1 dimensions, two key questions concerning the
properties of oscillons -- time-dependent, localized field configurations that
emerge during the deterministic evolution of $\phi^4$ models. First, we have
shown that initially asymmetric configurations evolve, for a wide range of
elliptic deformations, into symmetric oscillons states. Thus, oscillons are not
just particular to symmetric initial states. This result led us to propose that
oscillons are attractors in field configuration space, with a very deep
attractor basin, at least in 2+1 dimensions. Second, we have shown that
oscillons are stable against a wide range of asymmetric small perturbations. This
result was obtained by two distinct approaches, one solving the linearized equation
for the perturbations and the other by superimposing the perturbations on
the oscillons and evolving the perturbed configurations with the full equation
of motion. Clearly, both methods are restricted to the choice of initial
fluctuations. However, after an extensive search, we were unable to find any
unstable fluctuation with either approach. To the best of our knowledge,
this is the first dynamical investigation of the stability of explicitly time-dependent
scalar field configurations. We expect
that both of these results will carry on to 3+1 dimensions, although probably
the attractor basin will be shallower in this case.

These results suggest the importance time-dependent spatio-temporal
structures may have in a wide range of physical systems, from
condensed matter to early universe cosmology. Although we have restricted our
study to simple $\phi^4$ models, we expect, as suggested in Refs.
\cite{osc2,sornb}, that oscillons
will be present whenever there is a
bifurcation instability related to the negative curvature of the nonlinear
potential. Oscillons will emerge in a wide variety of dynamical systems,
possibly representing a bottleneck to equipartition of energy, thus delaying the
approach to equilibrium.

One possible arena for oscillons in early universe cosmology is during the
reheating supposed to occur after inflation. Oscillons may be thermally
nucleated with a probability proportional to $\exp[-E_{\rm osc}/T]$, where
$ E_{\rm osc}$ is the energy of the oscillon configuration. They will act as
``entropy sinks'', confining several degrees of freedom to an ordered state,
delaying the thermalization of the universe. Eventually, when they decay into
radiation, they will dump more entropy to the early universe, possibly changing
the final reheating temperature.

Finally, it would be interesting to compute the spectrum of quantum fluctuations
around oscillon states, to investigate their effect on oscillon stability.
Time-dependent bound states may have much to add to our knowledge of
nonperturbative quantum field theory, which has traditionally focused on
time-independent configurations, such as instantons.


\begin{acknowledgments}

The authors are indebted to the referee for pointing out a limitation of the
original stability analysis, which led to the present treatment.
A. B. Adib acknowledges Rafael
Howell for providing the numerical bounce solution for our stability analysis
and for many fruitful discussions, the Physics Department at UFC for the kind
hospitality during part of this work and Dartmouth College for financial
support. M. Gleiser was
partially supported by NSF grants PHY-0070554 and PHY-0099543, and by the ``Mr.
Tompkins Fund for Cosmology and Field Theory'' at Dartmouth. C. A. S. Almeida
was supported in part by CNPq (Brazil).

\end{acknowledgments}


\appendix*\section{Numerical method} The integration scheme adopted here is a
standard leapfrog algorithm, which  ensures second-order precision in time
\cite{numrec} (the spatial discretization used is fourth-order).
We have adopted the ``adiabatic damping method'' (or simply the damping method) of
Ref. \cite{sornb} together with Higdon's first-order boundary conditions
\cite{higdon}. Their combined use turned out to be very effective and of easy
implementation, allowing us to tackle this otherwise demanding numerical problem
with current workstations.

Put briefly, the leapfrog equations with the damping
method read:
\begin{eqnarray} \label{leapfrog}   \dot{\phi}^{n+1/2}_{i,j} & = &
\frac{(1- \eta_{i,j} \Delta t/2)\dot{\phi}^{n-1/2}_{i,j}
 + \Delta t [\nabla^2 \phi^{n}_{i,j} - V'(\phi^{n}_{i,j}) ]}{1+ \eta_{i,j}
\Delta t/2},\nonumber \\
  \phi^{n+1}_{i,j} & = & \phi^{n}_{i,j} + \Delta t
\dot{\phi}^{n+1/2}_{i,j},
\end{eqnarray}
where superscripts (subscripts) denote
temporal (spatial) indices, $\eta_{i,j}$ is the damping
function of Ref.
\cite{sornb} and $V'(\phi)$ is a first partial derivative of the potential with
respect to the field. The second spatial derivatives in the Laplacian are
discretized with a fourth-order scheme (to wit,
$\partial_{xx}\phi_{i,j} \approx
[ 16(\phi_{i+1,j} + \phi_{i-1,j}) - \phi_{i+2,j} - \phi_{i-2,j}
  - 30\phi_{i,j}]/12 \Delta x^2$ and analogously for $\partial_{yy}
\phi_{i,j}$),
which gives an energy conservation of one part in $10^3$
for $\Delta x=0.1$ and $\Delta t=0.06$
(and, of course, with $\eta_{i,j}\equiv 0$). A better energy conservation could
be obtained with smaller $\Delta x$ or $\Delta t$, but this comes with a high
price tag since, as remarked below, the computational time is inversely
proportional to both $\Delta x^2$ and $\Delta t$.
Despite this fact, with the
above parameters we were able to reproduce quite accurately the
results of
Gleiser and Sornborger \cite{sornb}. Even though the damping method is already a
major improvement over more naive
methods (such as huge lattices or even moving
boundary conditions), for the problem at
hand it is still demanding. As an
example, for small oscillons of radius
$R_0\approx 2.0$, the required lattice of
radial dimension $R
\approx 200$ adopted in Ref.
\cite{sornb} (and thus $L\approx
400$ in our square grid, where $L$ is the lattice edge),
would already demand a
total of $N \sim 10^{7}$ sites for $\Delta x=0.1$, as opposed
to the $N \sim
10^3$ used in the latter reference. Another aggravating fact comes from
the large integration times $\tau$ involved in such problems [notice that the
required computational time for this problem goes roughly as $\ctime \sim
(\tau/\Delta t) N = (\tau/\Delta t) (L/\Delta x)^2$]. We note in passing
that
there has been some effort to find a more natural and efficient discretization
for the $\phi^4$ theory which might reduce significantly the computational time
of such problems \cite{speight}. Motivated by this possibility, two of us have
recently investigated these lattices and have found that, unfortunately, they
are of limited practical use even for simple dynamical problems \cite{kinkdyn}.
It was seen, however, that if the above scheme is supplied with the boundary
conditions of Ref. \cite{higdon}:
\begin{equation}
\label{higdon}   \left.\left( \partial_t \pm \partial_{\alpha} \right) \phi
\right|_{\partial \Omega} =0,
\end{equation}
where $\alpha$ is either $x$ or $y$, then a significantly smaller lattice could
be used,
resulting in an energy error smaller than (or
equal to) the error due to numerical energy fluctuations. [These first-order
``absorbing boundary conditions'' were obtained for the rather simple
(linear) wave equation. We expect, however, that the damping introduced before
the boundaries could reduce the amplitude of the outgoing waves such that Eq.
(\ref{eqmotion}) is effectively linearized in that region, and thus that the
boundary condition (\ref{higdon}) becomes applicable].  With regard to the
example in the previous paragraph, we have found that the required lattice
with this ``mixed'' method needs only $L \approx 140$ (in contrast to the former
$L\approx 400$), such that $N$ (and thus $\ctime$) is roughly one order of
magnitude smaller than the previous one (this trend
is also found for greater $R_0$). For the sake of completeness, we quote
here the parameters of the
damping method used throughout our simulations (we use the same functional form
for $\eta(\rho)$ as Ref. \cite{sornb}): $k=0.005$ (damping constant), $\rho_0 =
10R_0$ (initial radius of the damping) and $\rho_\ell=50$ (damping length),
these latter two being
defined such that $L=2(\rho_0+\rho_\ell)$.

We expect that the method adopted here might be useful not only in higher
dimensional systems (the two methods above do not really make any dimensional
requirement), but also in other finite-domain problems not necessarily related
to oscillons.




\begin{thebibliography}{99}

\bibitem{rajaraman} R. Rajaraman, {\em
Solitons and Instantons} (North-Holland,  Amsterdam, 1987).

\bibitem{lee} T. D.
Lee and Y. Pang, Phys. Rep. {\bf 221}, 251 (1992).

\bibitem{coleman} S. Coleman, Nucl. Phys. {\bf B262}, 263 (1985).

\bibitem{nts-obs}  A. Kusenko, Nucl. Phys. B
(Proc. Suppl.) {\bf 62A/C},
248 (1998); A. Kusenko, M. Shaposhnikov, and P.
Tinyakov, Pisma Zh.Eksp.Teor.Fiz. {\bf 67}, 229 (1998); JETP Lett. {\bf 67}, 247
(1998); M. Axenides, E. Floratos, and A. Kehagias, Phys. Lett. B {\bf 444}, 190
(1998).

\bibitem{nontop-cosm} J. A. Frieman, G. B. Gelmini, M. Gleiser, and E. W.
Kolb, Phys. Rev. Lett. {\bf 60}, 2101 (1988); J. A. Frieman, A. V. Olinto,
M. Gleiser, and C. Alcock, Phys. Rev. D {\bf 40}, 3241 (1989); R. Friedberg, T. D.
Lee,  and Y. Pang, {\em ibid.} {\bf 35}, 3658 (1987), and references therein.
More recently, important consequences of NTS and $Q$ balls in cosmology   were
advanced by A. Kusenko and M. Shaposhnikov, Phys. Lett. B {\bf 418},
46 (1998);
A. Kusenko, V. Kuzmin, and M. Shaposhnikov, P. G. Tinyakov, Phys. Rev. Lett.
{\bf 80}, 3185 (1998); K. Enqvist and J. McDonald, Phys. Lett. B {\bf 425},
309 (1998). D.
Metaxas, Phys. Rev. D {\bf 63}, 083507 (2001);
A. Kusenko and P. J. Steinhardt,
Phys. Rev. Lett. {\bf 87}, 141301 (2001).


\bibitem{osc1}
M. Gleiser, Phys. Rev. D {\bf 49}, 2978 (1994).



\bibitem{osc2} E. J. Copeland, M. Gleiser and H.-R. Muller, Phys. Rev. D
{\bf 52},   1920 (1995). Oscillons at finite temperature
were investigated by M. Gleiser and R. M. Haas, Phys. Rev. D {\bf 54}, 1626 (1996).

\bibitem{breathers} S. Flach and C. R. Willis, Phys. Rep. {\bf 295}, 181 (1998);
D. K. Campbell, J. F. Schonfeld, and C. A. Wingate, Physica {\bf 9D}, 1 (1983).

\bibitem{sornb} M. Gleiser and A. Sornborger, Phys. Rev. E {\bf 62}, 1368 (2000).

\bibitem{chris} P. L. Christiansen, N. Gronbech-Jensen, P. S. Lomdahl, and
B. A. Malomed, Phys. Scripta {\bf 55}, 131 (1997).

\bibitem{honda} E. Honda and M. Choptuik, Phys. Rev. D {\bf 68}, 084037
(2002).

\bibitem{bogomolnyi} E. B. Bogomol'nyi, Sov. J. Nucl. Phys. {\bf 24}, 449 (1976).

\bibitem{calberto} F. S. A. Cavalcante, M. S. Cunha, and C. A. S. Almeida,
Phys. Lett. B {\bf 475}, 315. (2000).

\bibitem{jackiw} R. Jackiw, Rev. Mod. Phys. {\bf 49}, 681 (1977).

\bibitem{coleman2} S. Coleman, Phys. Rev. D {\bf 15}, 2929 (1977).

\bibitem{merkin} D. R. Merkin, {\em Introduction to the Theory of
Stability} (Springer, New York, 1997).

\bibitem{morrison} P. J. Morrison, Rev. Mod. Phys. {\bf 70}, 467 (1998).

\bibitem{lichtenberg} A. J. Lichtenberg, M. A. Lieberman, {\em Regular
and Chaotic Dynamics}, 2nd Ed. (Springer, New York, 1992).

\bibitem{derrick} G. H. Derrick, J. Math. Phys. {\bf 5}, 1252 (1964).

\bibitem{coleman3} S. Coleman, {\it Aspects of Symmetry}, Cambridge University Press
(Cambridge, UK 1985).

\bibitem{liu} X. Liu, L. Su, P. Ding, Int. J. Quant. Chem. {\bf 87}, 1 (2002).

\bibitem{numrec} W. H. Press, et al., {\em Numerical Recipes}, 2nd Ed.
(Cambridge  University Press. News York, 1992).

\bibitem{arfken} G. B. Arfken and H. J. Weber, {\em Mathematical Methods for
Physicists}, 4th Ed. (Academic Press, 1995).

\bibitem{higdon} R. L. Higdon, Math. Comp. {\bf 47} (1986), 437.

\bibitem{speight} J. M. Speight, Nonlinearity {\bf 10}, 1615 (1997). See also:
J. M. Speight and R. Ward, Nonlinearity {\bf 8}, 517 (1994); J. M. Speight,
Nonlinearity {\bf 12}, 1373 (1999).

\bibitem{kinkdyn} A. B. Adib and C. A. S. Almeida, Phys. Rev. E {\bf 64}, 037701
(2001).

\end{thebibliography}
\end{document}